\documentclass[12pt]{article}

\usepackage[tmargin=2cm,lmargin=2.5cm,bmargin=3cm,hmarginratio=1:1,headheight=65pt,headsep=1cm]{geometry} 
\usepackage[T1]{fontenc}
\usepackage[french,english]{babel}
\usepackage{color,graphicx}
\usepackage{amsmath,amssymb,amsfonts}
\usepackage{amsthm}
\usepackage{mathtools}
\usepackage{slashed}
\usepackage{sidecap} 
\usepackage{array}
\usepackage[indentafter]{titlesec}
\usepackage[utf8]{inputenc}
\usepackage{setspace}
\usepackage{perpage} 
\usepackage{fancyhdr} 
\usepackage{emptypage}
\usepackage{pifont} 
\usepackage{cite} 

\usepackage{hyperref} 

\usepackage{color}

\newcommand{\badat}{\begin{alignedat}}
\newcommand{\eadat}{\end{alignedat}}

\date{}
\begin{document}

\begin{titlepage}
  \thispagestyle{empty}
  \begin{center}  
\null 
\null
\null

{\LARGE\textbf{Cosmological horizons, Noether charges }}
\vskip0.12cm
{\LARGE\textbf{and entropy }}

\vskip1.2cm
   \centerline{Laura Donnay$^{a,b}$, Gaston Giribet$^{c}$}
\vskip1cm

{$^a$ Center for the Fundamental Laws of Nature, Harvard University}\\
{{\it 17 Oxford Street, Cambridge, MA 02138, USA.}}

{$^b$ Black Hole Initiative, Harvard University}\\
{{\it 20 Garden Street, Cambridge, MA 02138, USA.}}

{$^c$ Physics Department, University of Buenos Aires and IFIBA-CONICET}\\
{{\it Ciudad Universitaria, Pabell\'on 1, Buenos Aires, 1428, Argentina.}}

\end{center}

\vskip1cm

\begin{abstract}
It has recently been shown that, in the vicinity of their event horizons, black holes exhibit an infinite-dimensional symmetry. This symmetry captures relevant physical information about the black hole, and in particular about its thermodynamics. Here, we show that the same holds in a cosmological setup. More precisely, we show that around the de Sitter cosmological horizon, an infinite set of diffeomorphisms preserving a sensible set of boundary conditions emerges. These boundary conditions are similar to those considered by Price and Thorne in the context of the membrane paradigm, and permit to accommodate interesting gravity solutions. As for other boundary conditions considered previously, they are preserved by an infinite-dimensional asymptotic isometry algebra that includes supertranslations and two copies of the Virasoro algebra. This symmetry has associated a set of Noether charges that turn out to be conserved, finite, and integrable. We derive these charges explicitly using the covariant formalism and analyze their physical meaning by evaluating them to the case of de Sitter space. In this case, the zero-mode of the charge is found to account for the Gibbons-Hawking entropy of the cosmological horizon. We then consider a much more general set of solutions, including asymptotically de-Sitter black holes and asymptotically Taub-NUT-de Sitter black holes. 
\end{abstract}

\end{titlepage}

\section{Introduction} 

Observations indicate that we live in a dark energy dominated universe whose late time regime is well approximated by de Sitter space \cite{Perlmutter}. The accelerating fate of the universe implies the existence of a cosmological event horizon, a surface that surrounds any observer and that defines a point of no return: Distant comoving galaxies scape behind the cosmological horizon and information from there can not be retrieved, as it happens with matter falling into a black hole. In fact, black hole horizons and cosmological horizons share many mathematical and physical properties. In particular, cosmological horizons are also thermodynamical objects; they have temperature and entropy \cite{GH}. As for black holes, the entropy of cosmological horizons obeys the area law\footnote{We adopt the convention $\hbar =c =1$.}
\begin{equation}
S=\frac{\mathcal{A}}{4G}, \label{LaS} 
\end{equation}
with $\mathcal{A}$ being the area of the cosmological horizon, set by the Hubble scale. This fact has profound implications: In particular, it means that quantum gravity in de Sitter space, if consistent at all, demands the product of the cosmological constant $\Lambda$ and the Newton constant $G$ to take special, non-continuous values \cite{Witten}. It also raises the fundamental question about the actual reason for such special functional dependence with the area. In the last two decades, thanks to the holographic principle and its ramifications, we have gained insights about the meaning of the area law and of its generality. However, despite all these advances, a full understanding of it is still missing. The central question about the microscopic derivation of (\ref{LaS}) has only been answered in idealized examples. In the case of black hole horizons, the question can be addressed in the context of string theory, but the situation is quite different with de Sitter space, for which an embedding in string theory has shown to be notably elusive \cite{Vafa}. Probably, the most important advances in understanding the thermodynamics of de Sitter space came from the dS/CFT correspondence \cite{Strominger:2001pn}; but even in that context there are still open questions. 

Also in the case of black holes, which certainly offer a much more tractable scenario, there are important conceptual features of their thermodynamics yet to be understood. Such is the case of the information loss paradox, a longstanding problem that has recently been revisited. One of the recent attempts to address this problem is the idea of Hawking, Perry, and Strominger that the study of an intriguing infinite-dimensional symmetries --known as supertranslations-- that black holes exhibit in their near-horizon region \cite{Hawking:2016msc} could shed some light on the information hidden by the horizon (see also \cite{Hawking:2016sgy, Haco:2018ske} and references thereof). This study of infinite-dimensional symmetry is inspired by the recent application of similar ideas to the infrared physics in asymptotically flat space; see \cite{Strominger} and references therein. How much information is exactly encoded in this infinite-dimensional symmetry is still an open question. Nevertheless, the existence of such a large symmetry in the vicinity of event horizons is interesting on its own right, and the question remains as to what extent the thermodynamical properties of black holes can be captured by such a mathematical structure. In a more general context, it is interesting to investigate further the connection between the thermodynamical variables of event horizons, their infinite symmetries, and the associate conserved charges. The aim of this paper is precisely to extend to cosmological horizons the study of infinite-dimensional symmetries which have recently been successfully applied to black hole event horizons, and then apply it to investigate their thermodynamics\footnote{Infinite-dimensional symmetries in the context of cosmological models were also considered in \cite{Hotta1, Hotta2, Hotta25, Hotta26, Hotta3, Hotta4}.}. In order to do this in an efficient way, we will introduce a special set of asymptotic boundary conditions at the horizon, which are conveniently expressed in terms of Gaussian type coordinates. This will facilitate notably the task of finding the explicit change of coordinates needed to write the physically relevant solutions in the suitable near-horizon form. The discussion will be organized as follows: In Section II, we will review the main properties of de Sitter space and of its cosmological horizon. In Section III, we will derive the infinite-dimensional symmetry that emerges in the vicinity of the cosmological horizon and the Noether charges associated to it. In Section IV, we will apply these charges to de Sitter space and show that the zero-mode of the charge reproduces the Gibbons-Hawking entropy of the cosmological horizon. We will then extend the analysis to a much more general set of solutions: We will consider the case of asymptotically de Sitter and asymptotically Taub-NUT-de Sitter spaces, including black holes, for which we also study the infinite symmetries that appear both close to their black hole horizons and to their cosmological horizons. Section V contains our conclusions.  

\section{de Sitter space} 

De Sitter space is the maximally symmetric solution to Einstein equations with positive cosmological constant. It can be defined by a hyperbola embedded in (4+1)-dimensional Minkowski space: On $\mathbb{R}^{4,1}$, we define the metric
\begin{equation}
d{s}_5^2=-dX_0^2+dX_1^2+dX_2^2+dX_3^2+dX_4^2 \ ,\label{ds5}
\end{equation}
and impose the constraint
\begin{equation}
 X_1^2+X_2^2+X_3^2+X_4^2-X^2_0=\ell^2,\label{ell}
\end{equation}
where $\ell \in\mathbb{R} $ is the de Sitter radius. This defines a (3+1)-dimensional space with isometry group $SO(4,1)$, i.e. the remaining Lorentz part of the Poincar\'e isometry group of (\ref{ds5}), after the constraint (\ref{ell}) has broken the affine part of it. The space obtained by these means is non-compact and has positive, constant curvature; namely
\begin{equation}
R_{\mu\alpha\nu\beta}=\frac{1}{\ell^2} (g_{\mu\nu}g_{\alpha\beta}-g_{\mu\beta}g_{\nu\alpha}) \ ,
\end{equation}
which in particular implies the Einstein space condition $R_{\mu\nu}=\Lambda g_{\mu\nu}$ with cosmological constant $\Lambda=3/\ell^2$. 

In the so-called flat slicing, the metric of de Sitter space takes the form
\begin{equation}
ds^2= -d\tau^2 + e^{2\sqrt{\Lambda/3}\tau } (dx_1^2+dx_2^2+dx_3^2),\label{cosmo}
\end{equation}
which follows from the embedding
\begin{equation}
\badat{2}
&X_0=\ell\sinh(\tau /\ell) +\frac{\hat{r}^2}{2\ell }e^{\tau/\ell} ,   \ X_4=\ell\cosh(\tau /\ell) -\frac{\hat{r}^2}{2\ell }e^{\tau/\ell} ,\\
& X_1=e^{\tau /\ell } x_1   , \ \ \ X_2=e^{\tau /\ell } x_2  , \ \ \ X_3=e^{\tau /\ell } x_3  ,  \eadat
\end{equation}
with $\hat{r}^2=x_1^2+x_2^2+x_3^2$. This coordinate system only covers the region $X_0+X_4>0$ of the geometry. In these coordinates, de Sitter space has a clear cosmological interpretation: It describes an exponentially expanding isotropic, homogeneous universe with Hubble constant $H\equiv \ell^{-1}$. The space presents a cosmological horizon at the characteristic distance $H^{-1}$. The causal structure of the space, and in particular the existence of the horizon, is clearly captured by the static coordinate system, which is defined by
\begin{equation}
\badat{2}
&X_0=\sqrt{\ell^2-\hat{r}^2}\sinh(t/\ell)  , \  X_4=\sqrt{\ell^2-\hat{r}^2}\cosh(t/\ell), \\ 
& X_1=\hat{r} \zeta_1  , \  \ \ X_2=\hat{r} \zeta_2  ,\ \ \ X_3=\hat{r} \zeta_3  ,
\eadat
\end{equation}
with $\zeta_1^2+\zeta_2^2+\zeta_3^2=1$. In these coordinates the metric reads
\begin{equation}
ds^2 = (H^2\hat{r}^2-1)dt^2+(1-H^2\hat{r}^2)^{-1}d\hat{r}^2+\hat{r}^2\gamma_{AB}dz^Adz^B, \label{deSitter}
\end{equation}
where $\gamma_{AB}$ is the metric of the unit 2-sphere ($A,B= 1,2 $); $\zeta _i = \zeta_i (z^1,z^2)$ with $i=1,2,3$ give a standard embedding of $S^2$. We clearly observe from (\ref{deSitter}) that the event cosmological horizon appears at the constant radius $\hat{r}=H^{-1}$. The static patch corresponds to the region $\hat{r}^2\leq H^{-2}$; see Figure 1. In this region, the Killing vector $\partial_t$ is timelike. The time-dependent picture is recovered in the region $\hat{r}^2>H^{-2}$; in the large $\hat{r}$ limit this is given by $\hat{r}\simeq e^{\tau/\ell}+ ...$ where the ellipsis stand for terms that are subleading in the large $\tau$ limit. 
\begin{figure}
\ \ \ \ \ \ \ \ \ \ \ \ \ \ \  \ \ \ \ \ \ \ \ \ \ \ \ \ \ \ \ \ \ \ \includegraphics[width=2.8in]{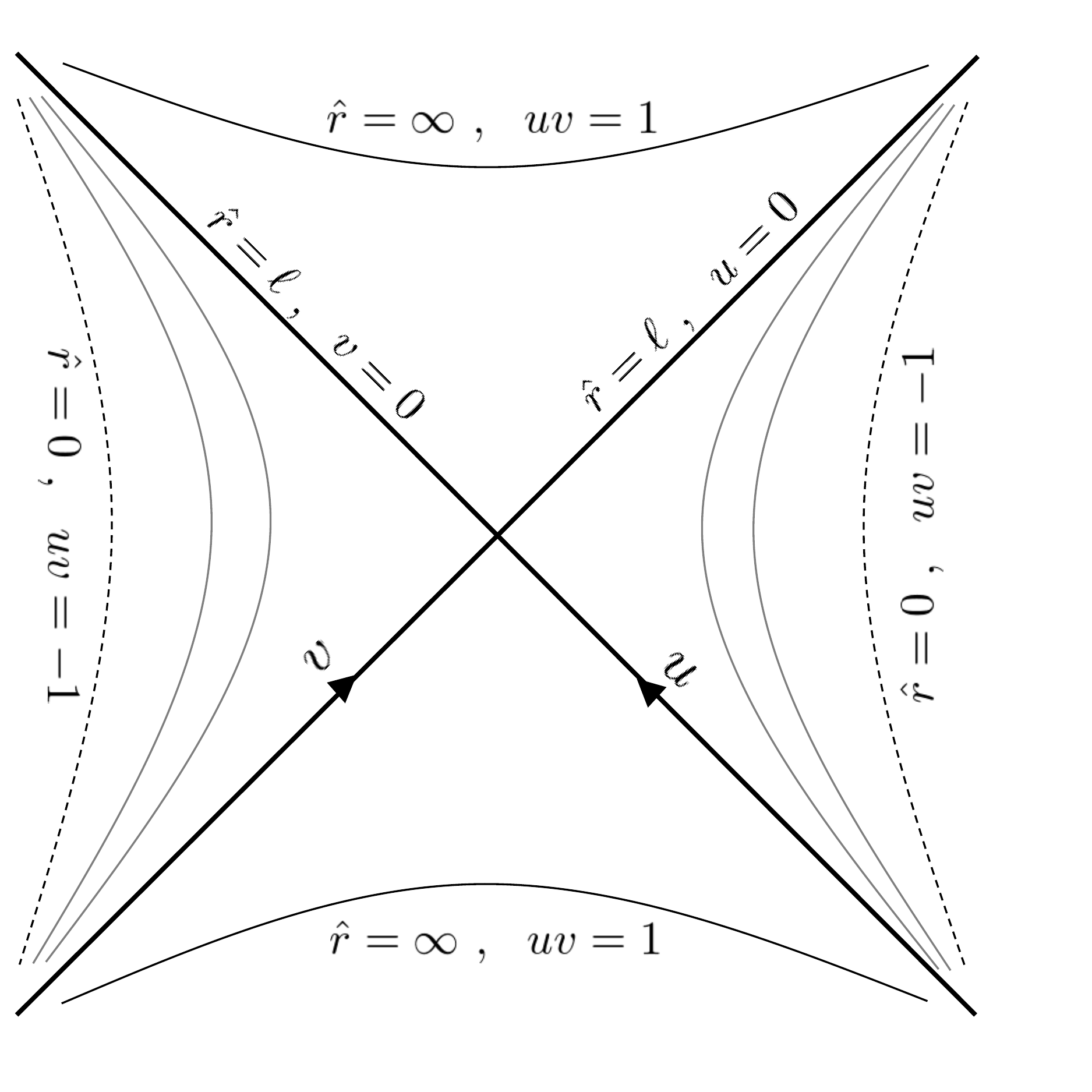}
\caption{Diagram of de Sitter space in Kruskal coordinates, defined by $\hat{r}=\ell (1+uv)(1-uv)$ and $uv=-e^{-2\tau /\ell }$. These coordinates permit to see the maximal extension of the geometry, which shows the existence of two pair of causally disconnected static patches separated by a past and a future event horizon.}
\label{Figure1}
\end{figure}

\section{Infinite-dimensional symmetry} 

\subsection{Horizon boundary conditions}

Now, let us analyze the region near the cosmological horizon and derive the infinite-dimensional symmetries that appear there. We can study this by prescribing a set of boundary conditions that employs Gaussian type coordinates similar to those considered in the context of the membrane paradigm \cite{Price:1986yy}. These coordinates, which are different from those recently considered to study black hole horizon symmetries in \cite{Donnay1, Donnay2}, have the advantage that the explicit change of coordinates from metrics in Boyer–Lindquist form can be easily written down \cite{Price:1986yy}. In particular, this is well suited to study the cosmological case: Consider the following near-horizon boundary conditions
\begin{eqnarray}
g_{tt}&=&-\kappa^2 r^2 + \mathcal{O}(r^4) , \ \ g_{tr} = \mathcal{O}(r^3) , \nonumber \\
g_{r r}&=&1 + \mathcal{O}(r^2), \ \ g_{A B}=\Omega_{AB}(z^C)+ \mathcal{O}(r^2) ,  \label{eq:2} \\
g_{r A}&=&\mathcal{O}(r^3), \quad g_{A t}=\kappa N_A(z^B) r^2 + \mathcal{O}(r^4) ,   \nonumber
\end{eqnarray}
where $\Omega_{AB}$ and $N_A$ are arbitrary functions of the angular coordinates $z^A$, $A,B, C=1,2$, and where $\mathcal{O}(r^n)$ stand for $z^A$-dependent functions that near $r= 0$ damp off as fast as $r^{\eta }$ with $\eta \geq n$. The horizon here is located at $r=0$. When accommodating de Sitter metric in the boundary conditions (\ref{eq:2}), one identifies $\kappa $ with the surface gravity at the horizon. Therefore, if one considers an isolated horizon, one has the condition $\kappa = const$.

Boundary conditions (\ref{eq:2}) are similar to those introduced by Price and Thorne in \cite{Price:1986yy} in the context of the membrane paradigm. These coordinates suffice to express the spacetime metric in a canonical form near the horizon, provided the horizon evolves slowly. In (\ref{eq:2}), time coordinate $t$ coincides with the universal time coordinate, and for de Sitter it is the time of the static patch close to the future horizon. This is different from the null coordinate $v$ of the boundary conditions considered in \cite{Donnay1, Donnay2} and which we review in section \ref{sec:otherbc}. Notice that the dependence of the lapse function $g_{tt}$ with the surface gravity $\kappa $ in (\ref{eq:2}) is quadratic, in contrast to the linear dependence $g_{vv}\propto \kappa $ of the the boundary conditions of \cite{Donnay1, Donnay2}; see also (\ref{eq:2314}) below. The boundary conditions  of \cite{Donnay1, Donnay2} also differ from (\ref{eq:2}) in that in the latter we are not fixing any gauge condition, allowing all the components of the metric to acquire subleading falling-off contributions, as in \cite{Price:1986yy}. One may still wonder about the connection between (\ref{eq:2}) and the boundary conditions of \cite{Donnay1, Donnay2}. For instance, one might think that both coordinate systems can be related by defining the tortoise coordinate  $r^*=\int^r g_{rr}d{r}$ together with the null coordinate $v=t+r^*$. However, although in the case of static spacetimes such change of coordinates may suffice to translate one system into the other, it is not generically the case: For example, in the case of Taub-NUT spacetime, which we will discuss later, the presence of an off-diagonal term in (\ref{eq:2}) under such a naive change of coordinates translates into a component of the metric that violates\footnote{In the notation of \cite{Donnay1,Donnay2}, it would result in a component $g_{A\rho }\simeq \mathcal{O}(\rho^0)$, where $\rho =0$ represents the horizon location. For details about the coordinate system of \cite{Donnay1,Donnay2} see section 4.2.} the boundary conditions of \cite{Donnay1,Donnay2}. The change of coordinates relating both systems is therefore more involved. Coordinates (\ref{eq:2}) certainly facilitate to express several relevant physical solutions in their near-horizon form.

\subsection{Asymptotic Killing vectors}

The set of metrics (\ref{eq:2}) is preserved by diffeomorphisms generated by vectors $\xi^{\mu}\partial_{\mu}$ obeying
\begin{equation}\label{eq:3}
\badat{3}
&\xi^{t}&=&T(z^A)+\mathcal{O}(r^4), \\ 
&\xi^{r}&=& \mathcal{O}(r^4), \\
&\xi^{A}&=&Y^A(z^C) + \mathcal{O}(r^4).
\eadat
\end{equation}
That is, an asymptotic Killing vector $\xi $ is defined by providing a set of three real functions $T(z^A)$, $Y^A(z^B)$ ($A, B =1,2$). This generates the asymptotic isometry group at the horizon which act on the metric functions in the following way:
\begin{equation}
\badat{2}\label{iop}
&\delta \Omega_{AB} &=& \mathcal{L}_Y \Omega_{AB},  \\
&\delta N_{A} &=& \mathcal{L}_Y N_A -\kappa \partial_A T,
\eadat
\end{equation}
together with $\delta \kappa = 0$. That is, functions $\Omega_{AB}$ and $N_A$, which are the leading terms in the small $r$ expansion of components $g_{AB}$ and $g_{At}$ respectively, have variations that depend on the Lie derivatives with respect to the 2-dimensional vectors $Y^A$ on the constant-$t$ slices of the horizon. In the case one considers on the horizon the conformal gauge $\Omega_{AB}\equiv \Omega \gamma_{AB}$, with $\Omega (z^A)$ being an arbitrary function of $z^A$ and $\gamma_{AB}$ being the constant curvature metric on the unit 2-sphere, the first equation of (\ref{iop}) reduces to
\begin{equation}
\delta \Omega = D_A (\Omega Y^A),
\end{equation}
with $D_A$ being the covariant derivative defined by $\gamma_{AB}$. In this case, $Y^A$ turn out to be conformal Killing vectors on the 2-sphere, and can be represented by holomorphic functions $Y^z(z)$ and anti-holomorphic functions $Y^{\bar{z}}(\bar{z})$.

Function $T(z^A)$, being an arbitrary function defined on the constant-$t$ foliation of the horizon, represents a supertranslation in the time direction; that is, a translation that a priori acts differently in each point $(z^1,z^2)$. Analogously, $Y^A(z^B)$ can be thought of as generating superrotations at the horizon. 

\subsection{Noether charges}

The asymptotic Killing vectors $\xi $ form an infinite-dimensional Lie algebra, which is defined by the modified Lie bracket introduced in \cite{Barnich3}; see \cite{Donnay2} for an explicit construction in the case of event horizons. Here, we will directly turn to the algebraic structure formed by the Noether charges associated to vector fields \eqref{eq:3}. By resorting to the covariant formalism \cite{Barnich:2001jy} one can derive these charges, which are found to be (see \cite{Donnay2} for details on this kind of computation):
\begin{equation}
Q[T,Y^A]=\frac{1}{16\pi G} \int d^2z \sqrt{\gamma }\,\Omega \left( 2T\kappa  -  Y^A N_A   \right)+Q_0 \ ,
\label{eq:4}
\end{equation}
where $\gamma\equiv\det {\gamma}_{AB}$, and where the integral is over the constant-$t$ section of the horizon. $Q_0$ is an arbitrary integration constant that sets the value of the charges for the background geometry. Notice that expression (\ref{eq:4}) is the integrated form of the charge. The covariant formalism rather gives the definition of the functional variation of the charge $\delta Q[\xi ]$, and it is not generally guaranteed that such expression is actually integrable. In fact, if one relaxes the isolated horizon condition and allows, for example, the function $\Omega_{AB}$ to depend on time, then $\delta Q[\xi ]$ turns out to be non-integrable. One can thus appreciate the fact that one could arrive to the integrated expression (\ref{eq:4}), which is moreover conserved and, provided the constant-$t$ foliations of the horizon is compact, finite. 

In order to express the charge algebra formed by the charges (\ref{eq:4}), it is convenient to consider complex coordinates $z,\bar{z}$, the conformal gauge $\Omega_{z\bar{z}}=\Omega(z,\bar{z}) \gamma_{z\bar{z}}$, and the mode expansion $T(z,\bar{z})=\sum_{m,n}T_{(m,n)}z^{m}\bar{z}^n$, $Y^z(z)=\sum_{m}Y^z_{m}z^{m}$, ${Y}^{\bar z}(\bar{z})=\sum_{m}{Y}^{\bar z}_{m}\bar{z}^m$, $m,n \in \mathbb{Z}$. As explained above, consistency demands $\partial_{\bar{z}}Y^z=\partial_zY^{\bar{z}}=0$. For example, local supertranslations are generated by $\xi = T_{(m,n)}z^{m}\bar{z}^n\partial_t $, with the standard (rigid) supertranslation in $t$ corresponding to $m=n=0$. Superrotations are generated by $Y_m$ and $\bar{Y}_m$. If one defines the supertranslation charge 
\begin{eqnarray}
\mathcal{T}_{(m,n)} &\equiv & Q[T_{(m,n)},Y^A=0]  \ ,\label{LaT}
\end{eqnarray}
and the superrotation charges
\begin{equation}
\badat{2}
\mathcal{Y}_{m}&\equiv & Q[T=0,Y^A=\delta^{A}_{z}Y^z_{m}],  \label{LaY}\\
\bar{\mathcal{Y}}_{m}&\equiv & Q[T=0,Y^A=\delta^{A}_{\bar{z}}Y^{\bar{z}}_{m}],
\eadat
\end{equation}
one has an appropriate basis. In this basis, defining the Poisson bracket as $\{Q [\xi_1], Q [\xi_2]\}=\delta_{\xi_1}Q [\xi_2]$, we find the following algebra:
\begin{equation}\label{virasoro}
\badat{3}
&\{{\mathcal{Y}}_m,{\mathcal{Y}}_n\}=(m-n)\mathcal{Y}_{m+n} \ , \\
&\{\bar{\mathcal{Y}}_m , \bar{\mathcal{Y}}_n\}=(m-n)\bar{\mathcal{Y}}_{m+n} \ ,\\
&\{\mathcal{Y}_m,\bar{\mathcal{Y}}_n\}=0 \ ,
\eadat
\end{equation}
together with
\begin{equation}
\badat{3}
&\{{\mathcal{Y}}_k,\mathcal{T}_{(m,n)}\}=-m \mathcal{T}_{(m+k,n)} \ , \\
&\{\bar{\mathcal{Y}}_k,\mathcal{T}_{(m,n)}\}=-n \mathcal{T}_{(m,n+k)} \ ,
\eadat
\end{equation}
and
\begin{eqnarray}
&&\{\mathcal{T}_{(k,l)},\mathcal{T}_{(m,n)}\}=0.\label{st}
\end{eqnarray}
While (\ref{LaT}) are the charges associated to the local generalization of translation $\partial_t $, (\ref{LaY}) are the charges associated to the local generalization of rotations $\partial_z - \partial_{\bar{z}}$. Together, these charges form an algebra that is the semi-direct sum of two copies of the Virasoro algebra (\ref{virasoro}) and supertranslation (\ref{st}). In the following section, we will evaluate these charges to the case of de Sitter cosmological horizon. This will allow us to investigate the physical meaning of (\ref{eq:4}). In particular, the charge $\mathcal{T}_{(0,0)}$, being related to a rigid translation in $t$, has to correspond to an {\it energy measure}. We will see below what physical quantity such energy corresponds to, and so for different types of solutions exhibiting event horizons.

\section{Cosmological horizons}

\subsection{de Sitter cosmological horizon}


A first interesting application of the boundary conditions (\ref{eq:2}) is the computation of charges associated to the de Sitter cosmological horizon, which is the simplest example of a more complete family of solutions we will consider later. In static coordinates, de Sitter is described by (\ref{deSitter}) and the cosmological horizon is located at $\hat{r}=H^{-1}\equiv \hat{r}_{++}$. Defining the inverse radial coordinate
\begin{equation}
r^2=\frac{1-H^2\hat{r}^2}{H^2}=\hat{r}_{++}^2 -\hat{r}^2,
\end{equation}
the metric (\ref{deSitter}) is easily accommodated in the boundary conditions (\ref{eq:2}) with functions 
\begin{equation}
\kappa = H  , \ \ \ \ \ \ \Omega_{AB}  = H^{-2}\gamma_{AB} ,  \ \ \ \ \ \ N_A=0. 
\end{equation}
Then, evaluating the charge (\ref{eq:4}) on de Sitter space yields vanishing charges except for the zero-mode $\mathcal T_{(0,0)}\equiv Q[T=1, Y^A=0]$ for which
\begin{equation}\label{TdS}
\mathcal{T}_{(0,0)} = \frac{H^{-1}}{8\pi G} \int d^2z \sqrt{\gamma }.
\end{equation}
This charge has a clear interpretation: Notice that $\kappa $ satisfies the equation $\chi^{\mu}\nabla_{\mu}\chi_{\nu}=\kappa \chi_{\nu}$, with $\chi^{\mu}$ being the components of the timelike Killing vector that generates the rigid translation in the coordinate $t$ and has norm one at the horizon. This confirms that $\kappa =H$ actually coincides with the surface gravity at the horizon, and therefore it gives the Gibbons-Hawking temperature of the cosmological horizon; namely,  
\begin{equation}
T=\frac{\kappa }{2\pi }= \frac{1}{2\pi}\sqrt{\frac{\Lambda}{3 }}.\label{TGH}
\end{equation}
On the other hand, the area of the horizon is given by 
\begin{equation}
\mathcal{A}=H^{-2}\int d^2z\sqrt{\gamma }. 
\end{equation}
Notice that this is valid for arbitrary horizon topology, regardless $\Omega_{AB}$ is the metric of a closed manifold, locally flat or locally of constant curvature.  

The zero-mode \eqref{TdS} can be rewritten in terms of these horizon quantities as
\begin{equation}
\mathcal{T}_{(0,0)}=\frac{\kappa}{2\pi } \frac{\mathcal{A}}{4G} = TS.\label{resultado}
\end{equation}
Therefore, since the boundary conditions (\ref{eq:2}) are defined for fixed $\kappa $, what the zero-mode of the supertranslation charge is actually computing is the Gibbons-Hawking entropy (\ref{LaS}) associated to the cosmological horizon (times the temperature). This simple example thus shows the physical meaning of the charge (\ref{eq:4}): the latter can be considered as a generalization of Wald's definition of entropy as a Noether charge on the horizon \cite{Wald:1993nt}, but now in the cosmological setup.
Notice that, in the case of a horizon with spherical topology, this result can be equivalently expressed as
\begin{equation}
\mathcal{T}_{(0,0)}=\frac{\Lambda }{6G}\hat{r}^3_{++} ,\label{resultadoX}
\end{equation}
which will turn out to be a convenient form in order to compare with more general solutions we will discuss in the next sections. 
 Notice also that $\mathcal{T}_{(0,0)}$ commutes with the charges $\mathcal{T}_{(m,n)}$, $\mathcal{Y}_n$, and $\bar{\mathcal{Y}}_n$; that is, those charges do not change the horizon entropy. 

\subsubsection{Comparison with other boundary conditions}
\label{sec:otherbc}
For sake of comparison, let us now consider the horizon boundary conditions of \cite{Donnay1, Donnay2} and apply them to the cosmological case: Close to a horizon located at $\rho=0$, consider the following asymptotic conditions
\begin{eqnarray}
g_{vv}&=&-2\kappa \rho + \mathcal{O}(\rho^2) ,  \nonumber \\
g_{A B}&=&\Omega_{AB}(z^C)+ \lambda_{AB}(z^C)\rho + \mathcal{O}(\rho^2) ,  \label{eq:2314} \\
g_{A v}&=&\theta_A(z^B) \rho + \mathcal{O}(\rho^2) ,   \nonumber
\end{eqnarray}
together with the gauge fixing conditions $g_{v\rho }=g_{\rho \rho}=0$ and $g_{v\rho}=1$; where $\Omega_{AB}$, $\lambda_{AB}$ and $\theta_A$ are arbitrary functions of the angular coordinates $z^A$, $A,B,C=1,2$. Null coordinate $v$ is the advanced time at the horizon and $\rho$ measures the radial separation from the horizon. 

In order to compare the computation performed with the boundary conditions (\ref{eq:2}), take de Sitter space in static coordinates (\ref{deSitter}) and perform the change of coordinates
\begin{equation}
\rho = \frac{1-H\hat{r}}{H}=\hat{r}_{++}-\hat{r} \ , \ \ \ dv= dt + \frac{d\rho }{2H\rho -H^2\rho^2} = dt - \frac{d\hat{r} }{1-{\hat{r}^2}/{\hat{r}_{++}^2}}.
\end{equation}
This yields the form
\begin{equation}
ds^2 = -2H\rho \ dv^2 +2dvd\rho +\gamma_{AB}(H^{-2}-2H^{-1}\rho)dz^Adz^B+\Delta g_{ij}dz^idz^j,\label{TYU55}
\end{equation}
where $\Delta g_{ij}\simeq \mathcal{O}(\rho^2)$ with $i,j\in \{v,A\}$. Then, we observe that (\ref{TYU55}) satisfies the boundary conditions (\ref{eq:2314}) with
\begin{equation}
\kappa = H \ , \ \ \ \Omega_{AB}=H^{-2}\gamma_{AB} \ , \ \ \ \lambda_{AB}=-2H^{-1}\gamma_{AB} \ , \ \ \ \theta_A = 0.
\end{equation}
These are the boundary conditions considered in \cite{Donnay1, Donnay2}, which are preserved by asymptotic Killing vectors of the form 
\begin{equation}
\xi = f(z^A)\partial_v + Y^A(z^B)\partial_A + \mathcal{O}(\rho);
\end{equation}
see \cite{Donnay2} for details. This yields the conserved charges
\begin{equation}
Q[f,Y^A]=\frac{1}{16\pi G}\int d^2z\sqrt{\gamma }(2f\kappa -Y^A\theta_A) ,
\end{equation}
for which, again, $\mathcal T_{(0,0)}=Q[f=1,Y^A=0]$ gives the product of the entropy and the temperature of the cosmological horizon, in perfect agreement with (\ref{resultado}). This shows that, despite the boundary conditions being different, the physical result is the same. Nevertheless, we emphasize the fact that Gaussian boundary conditions (\ref{eq:2}) have the advantage of being well suited to analyse more general asymptotically de Sitter solutions, as we explicitly show below.

\subsection{Schwarzschild-de Sitter event horizons}

Let us now consider black holes in de Sitter space. The metric describing such objects in Boyer-Lindquist type coordinates is given by Carter stationary solution
\begin{eqnarray}
ds^2 &&=  \frac{H(\theta )\ \sin ^2 \theta }{\rho^2 (\hat{r},\theta )}\Big( \frac{a\ dt - (\hat{r}^2+a^2)\ d\phi}{H(0)} \Big)^2 -\frac{\hat{r}^2F(\hat{r})}{\rho^2 (\hat{r},\theta )}\Big( \frac{dt-a\sin^2 \theta \  d\phi }{H(0)} \Big)^2  +\nonumber \\
&&\ \ \ \ \rho^2 (\hat{r},\theta ) \Big( \frac{d\hat{r}^2}{\hat{r}^2F(\hat{r})} + \frac{d\theta ^2 }{H(\theta )}\Big), 
\label{OPPO}
\end{eqnarray}
where
\begin{eqnarray}
F(\hat{r})&=&1-\frac{2GM}{\hat{r}}+\frac{a^2}{\hat{r}^2}-\frac{\Lambda }{3} (\hat{r}^2+a^2), \nonumber \\
H(\theta )&=&1+\frac{\Lambda a^2}{3}\cos^2 \theta , \label{jol} \\
\rho^2 (\hat{r},\theta )&=& \hat{r}^2+a^2\cos ^2\theta , \nonumber
\end{eqnarray}
and where $t\in \mathbb{R}$, $\hat{r}\in \mathbb{R}_{\geq 0}$, $\phi \in [0,2\pi ]$, $\theta \in [0,\pi ]$. This solution describes a Kerr-de Sitter black hole; $M$ represents the mass of the black hole and $J=aM$ its angular momentum.  The Carter solution above is a particular case of the Demia\'{n}ski-Pleba\'{n}ski solution; the latter including, in particular, the Taub-NUT charge $N$, which is absent in (\ref{OPPO})-(\ref{jol}). We will analyse the effects of the Taub-NUT in section \ref{sec:4.3}.

\subsubsection{The black hole event horizon}

The rotation effects from the horizon point of view were already analyzed\footnote{The rotating black hole case in terms of the Gaussian type coordinates (\ref{eq:2}) can be worked out by looking at the Appendix C of reference \cite{Price:1986yy}.} in \cite{Donnay1, Donnay2}. Here, as we are mainly interested in the effects on the cosmological horizon, it will be sufficient to consider the static case $a=0$. In this case, one obtains the Schwarzschild-de Sitter metric
\begin{equation}
ds^2=-F(\hat{r})dt^2+\frac{d\hat{r}^2}{F(\hat{r})}+\hat{r}^2\gamma_{AB}dz^A dz^B,\label{blac}
\end{equation} 
where $F(\hat{r})=1-{2GM}/{\hat{r}}-(\Lambda /{3}){\hat{r}^2}$, and where we recall that $\gamma_{AB}$ stands for the metric of the unit sphere. It is worth noticing that, providing $M\neq 0$, the cosmological horizon is not longer located at $({3/\Lambda })^{1/2}$, but at a radius that depends on both $\Lambda $ and $M$. In other words, there is a gravitational back-reaction on the cosmological horizon due to the black hole mass. In order to find the new position of the horizon, one needs to solve the polynomial equation 
\begin{equation}
\Lambda \hat{r}^3 -3 \hat{r} +6 GM=0. 
\end{equation}
This equation has three roots, given by the formula 
\begin{equation}
\hat{r} = \Big(-{3GM}/{\Lambda}\Big)^{1/3} \Big((1+x)^{1/3}+(1-x)^{1/3}\Big),\label{loo}
\end{equation}
with $x^2=1-1/(3\Lambda GM)^2$. Notice that (\ref{loo}) actually gives three, in principle different values, each of them corresponding to the different cubic roots of $1+x$. Provided the cosmological constant takes values in the range $0<\Lambda <1/(3 GM)^2$, the three values of (\ref{loo}) are real and obey the hierarchy $\hat{r}_{++}>\hat{r}_+>0>\hat{r}_{--}$; we are also assuming $M>0$. The cosmological horizon corresponds to the largest value $\hat{r}_{++}$, while $\hat{r}_+$ is the radius of the black hole horizon. We can easily verify that in the limit $\Lambda \to 0$ one recovers the Schwarzschild radius $\hat{r}_+=2GM$, while in the limit $M\to 0$ the horizon is $\hat{r}_{++}=(3/\Lambda)^{1/2}$.

Now, we want to investigate whether black hole solutions (\ref{blac}), near their horizons $\hat{r}_+$ and $\hat{r}_{++}$, can be accommodated in the boundary conditions (\ref{eq:2}). In the case of the black hole horizon, this can actually be achieved by considering the new coordinate $r$ defined by
\begin{equation}
r^2= \frac{2(\hat{r}-\hat{r}_{+})}{k (\hat{r}_+,\hat{r}_{++})},  \label{ttt}
\end{equation}
where the function $k (\hat{r}_+,\hat{r}_{++})$ is given by
\begin{equation}
k (\hat{r}_+,\hat{r}_{++}) = \frac{\Lambda }{6}\frac{(\hat{r}_{++}-\hat{r}_+)(\hat{r}_+-\hat{r}_{--})}{\hat{r}_+}.\label{tttT}
\end{equation}
The black hole horizon is now located at $r=0$, and in terms of this coordinate, the near-horizon limit of metric (\ref{blac}) takes the form (\ref{eq:2}) with the functions
\begin{equation}
\kappa = k (\hat{r}_+,\hat{r}_{++}) , \ \ \ \ \ \Omega_{AB}  = \hat{r}_{+}^{2}\gamma_{AB} ,  \ \ \ \ \ N_A=0.\label{estr}
\end{equation}
The value $\kappa = k (\hat{r}_+,\hat{r}_{++})$ actually coincides with the surface gravity of the black hole event horizon. This means that (\ref{resultado}) holds with the Hawking temperature $T_+=k (\hat{r}_+,\hat{r}_{++})/(2\pi ) $ and Bekenstein-Hawking entropy $S=\pi r_+^2/G$. This reproduces the result of \cite{Donnay1}, now in presence of $\Lambda > 0$. 

\subsubsection{The cosmological horizon}

In the case of the cosmological horizon, this can be done in the same way by simply replacing $\hat{r}_{+}\leftrightarrow \hat{r}_{++}$ in the definition (\ref{ttt}) and in (\ref{estr}). More precisely, one finds  
\begin{equation}
\kappa = -k (\hat{r}_{++},\hat{r}_{+}) ,  \ \ \ \ \ \Omega_{AB}  = \hat{r}_{++}^{2}\gamma_{AB} ,   \ \ \ \ \ N_A=0,\label{estEr}
\end{equation}
where $\kappa $ now coincides the surface gravity of the cosmological horizon, giving the Gibbons-Hawking temperature $T_{++}=|k (\hat{r}_{++},\hat{r}_{+})|/(2\pi )$ and entropy $S=\pi r_{++}^2/G$; i.e.
\begin{equation}
\mathcal{T}_{(0,0)} = \frac{\Lambda }{6G}\hat{r}_{++}(\hat{r}_{++}-\hat{r}_{+})(\hat{r}_{++}-\hat{r}_{--})=TS.\label{estrellitas}
\end{equation}
In the limit $M\to 0$ one recovers (\ref{resultadoX}). Expression (\ref{estrellitas}) shows that, even in presence of the black hole, where one has $M$ as an additional, independent dimension-1 quantity, the charge $\mathcal{T}_{(0,0)}$ computes the product of the temperature and the entropy, and this holds both for the black hole horizon and for the cosmological horizon.

\subsubsection{The Nariai limit}

The Nariai limit of the Schwarzcshild-de Sitter black hole solution, in which $\hat{r}_+$ tends to $\hat{r}_{++}$, can also be studied in this context. This limit produces a degenerate horizon. As pointed out in \cite{Donnay2}, the case of degenerate horizons is peculiar in what regards to its near-horizon symmetries and thus it has to be treated separately; this can be done by using the results of reference \cite{Nuevo}. 

The Nariai solution is obtained by taking in metric (\ref{blac}) the particular mass value
\begin{equation}
M=\frac{1}{3G\sqrt{\Lambda }}.
\end{equation}
In this case, we have $\hat{r}_+ = \hat{r}_{++}=1/\sqrt{\Lambda }$, and $\hat{r}_{--}=-2/\sqrt{\Lambda }$. Defining coordinates
\begin{equation}
\rho = \hat{r}-\frac{1}{\sqrt{\Lambda }} \ , \ \ \ \ \ dv=dt-\frac{1+\sqrt{\Lambda }\rho }{1+\frac 13 \sqrt{\Lambda }\rho }\ \frac{d\rho }{\Lambda \rho^2} ,
\end{equation} 
which for small $\rho $ yields
\begin{equation}
t=v-\frac{1}{\Lambda }{\rho}^{-1} +\frac{2}{3\sqrt{\Lambda }}\log \rho -\frac{2}{9} \rho +\frac{\sqrt{\Lambda }}{27}\rho^2 + ... ,
\end{equation} 
the metric takes the form
\begin{equation}
ds^2=\Lambda \rho^2 \ dv^2 + 2dvd\rho +({\Lambda}^{-1} + 2\Lambda^{-1/2}\rho ) \gamma_{AB}dz^Adz^B+\Delta g_{ij}dz^idz^j\label{extremal}
\end{equation}
where $\rho = 0 $ corresponds to $\hat{r}=\hat{r}_+=\hat{r}_{++}$, and where $\Delta g_{ij}\simeq \mathcal{O}(\rho^2)$, $i,j\in \{v,A\} $. This is precisely the form of the metric studied in \cite{Nuevo} to work out the symmetries near to extremal horizons. Such asymptotic form for the metric is preserved by Killing vectors of the form \cite{Donnay2}
\begin{equation}
\xi = X(z^A)v\partial_v + Y^A(z^B)\partial_A + \mathcal{O}(\rho ).
\end{equation}
Evaluated on (\ref{extremal}), this symmetry has associated a conserved charge
\begin{equation}
\mathcal{X}_{(0,0)} = \frac{1}{8\pi G\Lambda }\int d^2z\sqrt{\gamma }  = \frac{1}{2\pi }\frac{\mathcal{A}}{4 G}.\label{laveinticin}
\end{equation}
This result is analogous to what has been obtained in \cite{Donnay2} for extremal black holes, where the charge $\mathcal{X}_{(0,0)}\equiv Q[X=1,Y^A=0]$ was shown to reproduce the Bekenstein-Hawking entropy $S={\mathcal{A}}/{(4 G)}$ multiplied by a simple prefactor $1/(2\pi )$ that can be identified as the temperature $T_L$
that appears in the Kerr/CFT analysis of the extremal Frolov-Thorne vacuum; see \cite{KerrCFT, Frolov}. (\ref{laveinticin}) is the cosmological analogue of that.

\subsection{Taub-NUT-de Sitter horizons}
\label{sec:4.3} 
We move now to consider a more general set of solutions, namely the generalization for $\Lambda~>~0$ of the Taub-Newman-Unti-Tamburino spacetime (hereafter referred to as Taub-NUT-de Sitter). In the large $\hat{r}$ limit this actually yields a different asymptotic. We will consider here the Schwarzschild-Taub-NUT-de Sitter solution, which can be thought of as a generalization of the Schwarzschild-de Sitter black hole with gravitomagnetic charge. This is also a solution to cosmological Einstein equations in vacuum. Its metric can be written as follows
\begin{eqnarray}
ds^2 = -\frac{\hat{r}^2F(\hat{r})}{\rho^2 (\hat{r})} (dt +2NG \cos \theta  d\phi )^2 + \frac{\rho^2 (\hat{r})}{\hat{r}^2F(\hat{r})}d\hat{r}^2+\rho^2 (\hat{r}) (d\theta ^2 + \sin^2\theta d\phi ^2), \label{err}
\end{eqnarray}
with
\begin{equation}
\badat{2}
&F(\hat{r})= 1-\frac{2MG}{\hat{r}}-\frac{N^2G^2}{\hat{r}^2}-\frac{\Lambda }{3} (\hat{r}^2+6N^2G^2-\frac{3N^4G^4}{\hat{r}^2}),\\
&\rho^2 (\hat{r}) = \hat{r}^2+N^2G^2 ,
\eadat
\end{equation}
where, as $M$, $N$ is an arbitrary integration constant\footnote{Metric (\ref{err}) solves Einstein equations for arbitrary $M$ and $N$. However, as we will discuss later, these two parameters have to obey special relations for the Euclidean solution to be well defined.}. The solution reduces to Schwarzschild-de Sitter solution when $N=0$.

\subsubsection{The Schwarzschild-Taub-NUT black hole horizon}

In order to analyze some features of the Taub-NUT black hole solution (\ref{err}), let us first consider the case of vanishing cosmological constant. Replacing $\Lambda =0 $ in (\ref{err}) and performing the shift 
\begin{equation}
t\to t-2NG\epsilon \phi ,\label{boost}
\end{equation} 
we find
\begin{eqnarray}
ds^2 = -U(\hat{r}) (dt +2NG (\cos \theta -\epsilon ) d\phi )^2 + \frac{d\hat{r}^2}{U(\hat{r})} +(N^2G^2+\hat{r}^2) (d\theta ^2 + \sin^2\theta d\phi ^2) 
\end{eqnarray}
with $U(\hat{r})= (\hat r^2-2MG\hat{r}-N^2G^2)/({N^2G^2+\hat{r}^2})$, where now $\hat{r}\in \mathbb{R}$, $\phi\in [0,2\pi ]$, $\theta \in [0,\pi ]$, and $\epsilon =\pm 1$. The change of coordinates (\ref{boost}) is needed to cure the wire-type singularity that the solution (\ref{err}) exhibits at $\theta =0$ and $\theta =\pi $. As shown by Misner, this can be done by considering two patches: $\epsilon =+1$ in the hemisphere $\theta <\pi /2$ and $\epsilon =-1$ in the hemisphere $\theta >\pi /2$. Due to the periodicity of $\phi $, (\ref{boost}) demands $t$ to be periodic too, with a period 
\begin{equation}
\beta=4\pi NG. 
\end{equation}
Taub-NUT space can be regarded as the gravitational analog of a magnetic monopole, and the periodicity in $t$ can be thought of as the gravitational analog of the Dirac quantization condition, with $N$ being the gravitomagnetic charge. The space constructed in this way turns out to be spatially homogeneous and of topology $\mathbb{R}\times S^3$. It has isometry $SO(3)$, and thus is spherically symmetric. It contains closed timelike curves, though. Provided $N\neq 0$, the surface $\hat{r} =0$ does not present curvature singularity, and so one can consider the entire range $\hat{r}\in \mathbb{R}$. The solution, however, does exhibit coordinate singularities at $\hat{r}=\hat{r}_{\pm}$, with 
\begin{equation}
\hat{r}_{\pm} = GM\pm G\sqrt{M^2+N^2},
\end{equation} 
being the two solutions of $U(\hat{r}_{\pm})=0$. The radii $\hat{r}_{\pm }$ can be considered as horizons. For $N=0$, the solution reduces to Schwarzschild metric, and in that case $\hat{r}\in \mathbb{R}_{\geq 0}$ with $\hat{r}_-=0$ and $\hat{r}_+=2GM$. For arbitrary values of $M$ and $N$, the function $U(\hat{r})$ behaves at large $\hat{r}$ like $U(\hat{r}) \simeq 1-2MG/\hat{r} + \mathcal{O}(N^2/\hat{r}^2)$, so that $M$ represents the mass of the solution. On the other hand, $N$ enters in the large $\hat{r}$ expansion as the charge contribution does in the Reissner-Nordstr\"{o}m solution, although it is of purely gravitational origin.  The fact that when $N\neq 0 $ the off-diagonal component $g_{\phi t}$ does not vanish at infinity implies that, strictly speaking, the solution with $\Lambda =0$ is not asymptotically flat, but it rather exhibits a special kind of asymptotic behavior characterized by the parameter $N$ and known as asymptotically locally flat (ALF). Provided $M $ is positive --as we will consider hereafter-- the solution has two horizons, located at $\hat{r}_{+ }>0$ and $\hat{r}_{- }<0$. Since $M=({\hat{r}_+ + \hat{r}_-})/({2G}) \geq 0$ and $N^2 = -\hat{r}_- \hat{r}_+ /G^2 \geq 0$, one can write the metric function $U(\hat{r})$ as follows
\begin{equation}
U(\hat{r})= \frac{(\hat{r}-\hat{r}_+)(\hat{r}-\hat{r}_-)}{\hat{r}^2-\hat{r}_-\hat{r}_+}.
\end{equation}
In terms of the horizon radii, the metric takes the form
\begin{eqnarray}
ds^2 &=& -U(\hat{r})dt^2 -4\sqrt{-\hat{r}_-\hat{r}_+}U(\hat{r})(\cos \theta -\epsilon ) dt d\phi +  4\hat{r}_-\hat{r}_+U(\hat{r})(\cos\theta -\epsilon )^2 d\phi^2\nonumber \\
&&+\frac{d\hat{r}^2}{U(\hat{r})}+ (\hat{r}^2-\hat{r}_-\hat{r}_+)(d\theta ^2 + \sin^2 \theta d\phi ^2 ).
\end{eqnarray}

We are now ready to study the near-horizon region of the Taub-NUT black hole: We define the near-horizon coordinate
\begin{equation}
r^2 = 4\hat{r}_+(\hat{r}-\hat{r}_+),
\end{equation}
so that $r=0$ corresponds to $\hat{r}=\hat{r}_+$. Then, close to the black hole event horizon we find
\begin{equation}
\badat{3}
&g_{tt}=-\frac{r^2}{4\hat{r}_+^2} + \mathcal{O}(r^4) , \ \ g_{tr} = 0 ,  \\
&g_{r r}=1 + \mathcal{O}(r^2), \ \ g_{A B}=\hat{r}_+(\hat{r}_+ - \hat{r}_-)\gamma_{AB}+ \mathcal{O}(r^2) ,  \\
&g_{r A}=0, \quad \ \ g_{A t}=\delta_A^{\phi} \frac{r^2\sqrt{-\hat{r}_+\hat{r}_-}}{2\hat{r}^2_+}(\epsilon -\cos \theta ) + \mathcal{O}(r^4) ,  
\eadat
\end{equation}
with $A,B\in \{\phi , \theta \}$. This obeys (\ref{eq:2}), with
\begin{equation}\label{re}
\kappa = \frac{1}{2\hat{r}_+} , \ \ \ \ \ \Omega_{AB}= \hat{r}_+(\hat{r}_+ - \hat{r}_-)\gamma_{AB}, \ \ \ \ \ N_A=\delta_A^{\phi }(\epsilon -\cos \theta ) {\sqrt{-{\hat{r}_-}/{\hat{r}_+}}}  . 
\end{equation}

From (\ref{re}), we find the charge
\begin{equation}
\mathcal{T}_{(0,0)} = \frac{(\hat{r}_+-\hat{r}_-)}{4 G} = \frac{1}{2}\sqrt{M^2+N^2} ,\label{FGH}
\end{equation}
which reduces to the result for the Schwarzschild solution found in \cite{Donnay1} when $N=0$, i.e. when $\hat{r}_{-}=0$. More precisely, when $N=0$ (\ref{FGH}) gives one half of the ADM mass, $\mathcal{T}_{(0,0)} =M/2$, which coincides with the product $TS$ through the Smarr formula. In the case $M=0$, on the other hand, (\ref{FGH}) gives one half of the NUT charge, $\mathcal{T}_{(0,0)} =N/2$. Notice that here we are writing (\ref{FGH}) in such a way that no special values of the parameters $M$ and $N$ were assumed. This is why one should not expect $\mathcal{T}_{(0,0)}$ to give in general the product of the entropy $S= \mathcal{A}/(4G)$ times the temperature $T=U'(\hat{r}_+)/(4\pi )=1/(4\pi \hat{r}_+)$ for arbitrary $\hat{r}_{\pm }$. Having this would demand a special relation between the parameters of the solution for cycles of the Euclidean Taub-NUT geometry to be uniquely defined; i.e. it would demand $T=\beta^{-1}=1/(4\pi N G)$ that yields the usual condition $r_+=NG$, in which case $\hat{r}_+-\hat{r}_-=2NG$ and $\mathcal{A}=4\pi \hat{r}_+(\hat{r}_+ - \hat{r}_-)=8\pi N^2G^2$. In fact, for these particular values (\ref{FGH}) takes the form
\begin{equation}
\mathcal{T}_{(0,0)}=\frac{N}{2} = \beta^{-1}\frac{\mathcal{A}}{4G},
\end{equation}
which gives the NUT charge and in this case coincides with the inverse of the Euclidean time period multiplied by one quarter of the area in Planck units. That is to say, the charge (\ref{FGH}) does reproduce the expected result once the adequate relation between the parameters is considered. We do not find necessary to repeat here the standard discussion\footnote{The thermodynamics of Taub-NUT and Taub-Bolt solutions including a cosmological constant has been discussed in \cite{HHP}, and a discussion of the de Sitter case can be found in \cite{Mann1, Mann2}.} on Taub-NUT and Taub-Bolt spaces and the consistency conditions for the real section of the Euclidean solutions with $\hat{r}_+\geq NG$ to be well defined. We rather prefer to write (\ref{FGH}) in such a way that the expression of the horizon charge $\mathcal{T}_{(0,0)}$ as a function of $\hat{r}_{\pm }$ tends to the result for the Schwarzschild black hole when $N$ goes to zero.

Notice that, in the case of the Taub-NUT (-de Sitter) black hole, unlike for the case of the Schwarzschild(-de Sitter) black hole, $N_A$ is different from zero. In the case of rotating black holes, the non-vanishing value of $N_A$ results in a contribution to the charge $Q[T=0,Y^A]$ that accounts for the black hole angular momentum \cite{Donnay1}; in the case of Schwarzschild-Taub-NUT solution, in contrast, we find
\begin{equation}
\mathcal{Y}_{0}-\bar{\mathcal{Y}}_{0}=Q[T=0,Y^A=\delta^A_{\phi }] = 0.
\end{equation}
That is, the charge associated to the Killing vector $\xi =Y^A\partial_A$ vanishes despite $N_A\neq 0$. This is due to the different values that $\epsilon $ takes in each hemisphere for the metric to be singularity free, what makes $\mathcal{Y}_{0}-\bar{\mathcal{Y}}_{0}$ to be proportional to the angular integral $\int_0^{\pi/2}d\theta \sin \theta (\cos \theta -1)+\int_{\pi /2}^{\pi }d\theta \sin \theta (\cos \theta +1) =0$. This is analogous to what happens with the horizon superrotation charge of dyons in Einstein-Maxwell theory \cite{Nuevo}. A way of interpreting this result is that $\mathcal{Y}_{0}-\bar{\mathcal{Y}}_{0}$ computes the Misner string contribution to the angular momentum when $|\epsilon |\neq 1$. 

\subsubsection{The Schwarzschild-Taub-NUT/Bolt-de Sitter cosmological horizon}

Now, let us go back to cosmology: consider the case $\Lambda >0$, for which the solution (\ref{err}) represents a Schwarzschild-Taub-NUT-de Sitter black hole. We can write
\begin{equation}
{\hat{r}^2F(\hat{r})} = -\frac{\Lambda}{3}\ {(\hat{r}-\hat{r}_{++})(\hat{r}-\hat{r}_{+})(\hat{r}-\hat{r}_{-})(\hat{r}-\hat{r}_{--})},
\end{equation}
where $\hat{r}_{++}$, $\hat{r}_{+}$, $\hat{r}_{-}$, and $\hat{r}_{--}$ are the four roots of $F(\hat{r})=0$. In particular, it implies $r_{++}r_{+}r_{-}r_{--}=3N^2G^2(1/\Lambda -N^2G^2)$. For a $\Lambda $-dependent range of the parameters $M$, $N$ these four roots are real and obey the hierarchy $\hat{r}_{++}>\hat{r}_{+}>0>\hat{r}_{-}>\hat{r}_{--}$, with $\hat{r}_{++}$ representing the cosmological horizon and $\hat{r}_{+}$ the black hole horizon. As in the case of the Schwarzschild-de Sitter black hole, in order to analyze the region near the cosmological horizon we define the coordinate 
\begin{equation}
r^2=\frac{2(\hat{r}-\hat{r}_{++})}{k(\hat{r}_{++},\hat{r}_+)},
\end{equation}
with 
\begin{equation}
k(\hat{r}_{++},\hat{r}_+) = \frac{\Lambda }{6}\frac{(\hat{r}_{+}-\hat{r}_{++})(\hat{r}_{++}-\hat{r}_{-})(\hat{r}_{++}-\hat{r}_{--})}{(\hat{r}^2_{++}+N^2G^2)},\label{54}
\end{equation}
such that $r=0$ when $\hat{r}=\hat{r}_{++}$. Notice that (\ref{54}) reduces to the value of $k(\hat{r}_{++},\hat{r}_+)$ for the Schwarzschild-de Sitter black hole when $N=0$, i.e. when $\hat{r}_-=0$. This yields the charge
\begin{equation}
\mathcal{T}_{(0,0)} = \frac{\Lambda }{6G} (\hat{r}_{++}-\hat{r}_{+}) (\hat{r}_{++}-\hat{r}_{-}) (\hat{r}_{++}-\hat{r}_{--}),\label{turqueli}
\end{equation}
which, as expected, agrees with the result for the cosmological horizon of the Schwarzschild-de Sitter black hole (\ref{estrellitas}) when $\hat{r}_-=0$. That is, expression (\ref{turqueli}) comes to generalize (\ref{resultadoX}) and (\ref{estrellitas}) which now appear as the particular cases $\hat{r}_{++}>\hat{r}_{+}=\hat{r}_{-}=0$ and $\hat{r}_{++}>\hat{r}_{+}\neq \hat{r}_{-}=0$, respectively. It also agrees, of course, with the result obtained in \cite{Donnay1} for the Schwarzschild black hole $\hat{r}_{++}=\infty > \hat{r}_+ \geq \hat{r}_- = 0$. 

\subsection{Other solutions}

It is actually possible to extend the analysis of horizon symmetries to other solutions with regular cosmological horizons. One such example is the McVittie solution of Einstein equations \cite{Mc}, which describes a black hole in an expanding universe. The universe in McVittie solution is described by a fluid with an energy density $\sigma $ that is constant at constant-time slices and with a pressure $p$ that acquires an inhomogeneous contribution due to the presence of the black hole. This solution has been recently revisited in \cite{Matt}, where it has been noticed that, provided a positive cosmological constant is present, the McVittie solution is regular everywhere on and outside the black hole event horizon and away from the big bang singularity. The metric can be casted in the form
\begin{equation}
ds^2 = - (F(\hat{r})-H^2(t)\hat{r}^2) dt^2  - \frac{2H(t)\hat{r}}{\sqrt{F(\hat{r})}} d\hat{r} dt +\frac{d\hat{r}^2}{F(\hat{r})} + \hat{r}^2(d\theta^2 + \sin^2\theta d\phi^2),
\end{equation}
where $F(\hat{r}) = 1-{2MG}/{\hat{r}}$, with $M$ being the mass of the black hole, and where $H(t)$ obeys the Friedmann equation $H^2(t)=8\pi G\sigma (t) /3$. This solution reduces to the Schwarzschild-de Sitter solution when $H$ is a positive constant, and it takes the Friedmann–Lema\^{\i}tre–Robertson–Walker (FLRW) form when $M=0$. 

As said, in the case $H(t=\infty ) > 0$ the McVittie metric describes a black hole whose horizon is regular and is embedded in an FLRW spacetime. Its causal structure is quite interesting, as it has a spacelike and inhomogeneous big bang singularity in the remote past. It also has a spacelike future infinity at large radial distance $\hat{r}$ and late time $t$. The surface $\hat{r} = 2MG$ is a curvature singularity at finite $t$. The solution also contains a null apparent horizon at $\hat{r} = \hat{r}_-$ and $t=\infty $, where $\hat{r}_-$ is the smaller positive root of the cubic, time-dependent equation $F(\hat{r}_{\pm})=H^2(t)\hat{r}_{\pm}^2$. Provided $H(t=\infty ) > 0$, the solution also exhibits a cosmological event horizon at the point $\hat{r} = \hat{r}_+(t=\infty )$ and $ t =\infty $; at $\hat{r}=\hat{r}_{\pm}(t)$ and $t$ finite, the solution exhibits an apparent horizon. Analyzing the near-horizon symmetries and the associated conserved charges for this type of solutions with the techniques discussed here\footnote{More precisely, consider the coordinate system defined in Eqs. (20)-(22) of \cite{Matt} and compare it with the system (\ref{eq:2314}) herein, identifying $v=\tau $ and $\rho=r-r_-$.}
 is also possible. The result is consistent with (\ref{turqueli}). Other solutions with apparent horizons can also be studied in the same way.

\section{Conclusions}

The results of this paper provide an extension to the cosmological setup of previous studies \cite{Donnay1, Donnay2} of infinite-dimensional symmetries in the vicinity of event horizons: We have shown that, in its proximity, cosmological de Sitter horizons exhibit an infinite-dimensional symmetry. The latter is the asymptotic isometry group that preserves a suitable set of boundary conditions adapted to study cosmological horizons, boundary conditions which happen to be similar to those considered by Price and Thorne in the context of the membrane paradigm. The corresponding isometry group is generated by an infinite-dimensional algebra, which is identified as the semi-direct sum of two copies of the Virasoro algebra (with vanishing central charge) and supertranslations. This symmetry has an associated Noether charge, which we have derived using the Barnich-Brandt covariant formalism. This charge is conserved, integrable, and finite, and we discussed its physical meaning by evaluating it on de Sitter space. The zero-mode of this charge was found to give the Gibbons-Hawking entropy of the cosmological horizon. We then studied a larger class of geometries, considering the case of Schwarzschild-Taub-NUT-de Sitter spaces: we found that their associated Noether charge, in region close to both the black hole horizon and the cosmological horizon, computes the entropy of these horizons.

The fact that symmetries at the cosmological horizon yield conserved charges associated to their thermodynamics is interesting on its own right. A much more interesting, although speculative question is whether these symmetries somehow encode information hidden behind the cosmological horizon. As pointed out in \cite{Susskind}, it is an interesting conceptual question whether the matter passing out through the cosmological horizon will leave behind some imprint in the same way that information that fell through a black hole reappears in a scrambled form in the Hawking radiation. It would be interesting to study whether the infinite-dimensional symmetries discussed here have something to say in this regard.

Another interesting application of the near-horizon symmetries of cosmological horizons is the connection of such symmetries with the membrane paradigm \cite{Wang} and, more specifically, with the Carroll ultra-relativistic symmetries that black hole horizons exhibit \cite{Penna, Lau}. To this respect, it is worth emphasizing that the boundary conditions we considered here are well suited to the membrane paradigm formalism \cite{Price:1986yy}.

\subsection*{Acknowledgments}

The authors are grateful to Hern\'{a}n Gonz\'{a}lez for collaboration at early stages of this project. LD acknowledges support from the Black Hole Initiative at Harvard University, which is
funded by a grant from the John Templeton Foundation.


\end{document}